\documentclass[a4paper]{jpconf}
\usepackage{graphicx}
\usepackage{textcomp}
\begin{document}
\title{High-power test results of a 3 GHz single-cell cavity}

\author{U. Amaldi$^{1}$, D. Bergesio$^{1}$, R. Bonomi$^{1,4}$, A. Degiovanni$^{1,2}$, M. Garlasch\'e$^{1,}$\footnote[4]{Now at CERN.}, P. Magagnin$^{1}$, S. Verd\'u-Andr\'es$^{1,3}$ and R. Wegner$^{4}$}

\address{$^{1}$ TERA Foundation, Via Puccini 11, 28100 Novara, Italia}
\address{$^{2}$ EPFL, Route Cantonale, 1015 Lausanne, Switzerland}
\address{$^{3}$ IFIC (CSIC-UV), Ed. Institutos de Investigaci\'on, C/ Catedr\'atico Jos\'e Beltr\'an 2, 46980 Paterna, Spain}
\address{$^{4}$ CERN, Route de Meyrin 385, 1211 Geneva, Switzerland}

\ead{Silvia.Verdu.Andres@cern.ch}

\begin{abstract} % 200 words, one paragraph, include results and conclusions
Compact, reliable and energetically efficient accelerators are required for the treatment of tumours with ions. TERA proposes the ``cyclinac'', composed of a fast-cycling cyclotron and high-frequency booster linac. The dimensions of the linac can be reduced if high accelerating gradients are used. TERA initiated a high-gradient test program to understand the operational limits of such structures. The program foresees the design, prototyping and high-power testing of several high-gradient structures operating at 3 and 5.7 GHz. The high-power tests of the 3 GHz single-cell cavity were completed in March 2012. The maximum break down rate (BDR) threshold measured for a peak electric surface field $E_{max}$ of 170~MV/m and a RF flat top pulse length of 2.5~\textmu s was $3\times10^{-6}$~bpp/m. 
\end{abstract}

\section{Motivation}
The hadrontherapy community demands compact, reliable and energetically efficient accelerators for tumour treatment with ions. CABOTO is a normal-conducting, high-frequency, fast-cycling linac designed by TERA which boosts the energy of the particles previously accelerated by a cyclotron~\cite{bib:Amaldi_NIMA10}. The length of CABOTO could be below 26 m if high gradients in the order of 35~MV/m on axis were used, which would lead to maximum electric fields in the order of 170~MV/m in the structure. 

The use of high-gradient structures for medical purposes requires to prove that the machine can operate at high gradients without compromising its reliability. Therefore, the study of high-gradient structure performances becomes essential to understand the possible use of this kind of structures in hadrontherapy. A collaboration between TERA and the CLIC - RF structure development group at CERN has been established to advance together in this field. Although TERA and CLIC are working on very different RF structures, both share the same operational limits: a 200~MV/m maximum electric field and a maximum $10^{-6}$~BDR (breakdown rate, the number of breakdowns per pulse per meter of an accelerating structure)~\cite{bib:Grudiev}. This corresponds to about one breakdown per treatment course, the acceptable breakdown rate for a medical accelerator.

In this framework, TERA initiated a high-gradient test program~\cite{bib:Degiovanni_NIM11} which consists in the design, prototyping and high-power tests of one 3 GHz single-cell cavity and three 5.7 GHz single-cell cavities. The main goal of this program is to understand the limitation at which the high-gradient CABOTO-like cavities can operate reliably. The program foresees the design, prototyping and high-power test of a multi-cell structure operating at the frequency which provided better performances. The high-power tests of the 3 GHz single-cell cavity were completed in March 2012 and are presented in the following, preceded by a brief description of the cavity design, prototyping and low power RF measurements. 

\section{3 GHz single-cell cavity design and fabrication}

\begin{figure}[ht]
\begin{center}
\includegraphics[width=15cm]{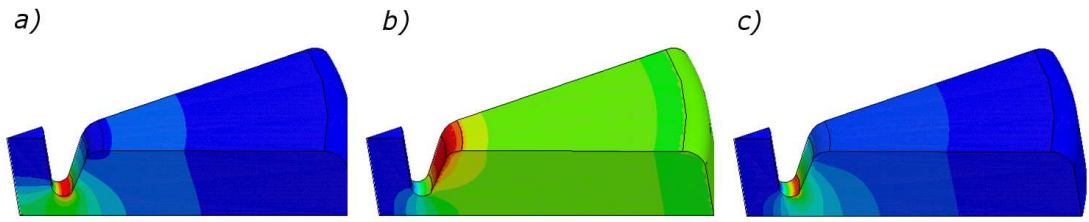}
\end{center}
\caption{\label{fig:fieldDistr} Distribution of a) electric field $E$, b) magnetic field $H$ and c) modified Poynting vector $S_{c}$, in the cell volume.}
\end{figure}

The cell geometry was similar to the low-energy cells of the high-gradient linacs designed by TERA, presenting nose cones that enhance the electric field along the beam axis. It was optimized to maximize the shunt impedance for a bore radius of 3.5~mm. The RF design of the cavity was based on 2D and 3D simulations, respectively done with Poisson Superfish and Ansoft HFSS. Fig.~\ref{fig:fieldDistr} shows the distribution of electric and magnetic field and modified Poynting vector $S_{c}$ in the cell volume. The maximum electric and magnetic field and the maximum modified Poynting vector for the designed cell are given in Table~\ref{tab:mainEM}. 

The power was provided to the cell by magnetic coupling. The cavity geometry was designed such that the structure was overcoupled. A movable short was placed at one end of the waveguide in order to match the waveguide feeder to the cell. Subsequently, the short was brazed at the position which led to critical coupling.

%The frequency shift due to thermal expansion was calculated to be -0.05~MHz/degree. The thermal resistance of the cavity is 0.035 degrees/W. Thus a total detuning of around -2 MHz was expected when the coolant inlet temperature is 15 degrees above ambient temperature and the cavity is fed with 350 W average power. 
The frequency shift due to thermal expansion was calculated to be -0.05~MHz/degree. The thermal resistance of the cavity is 0.035~degrees/W. 
%Thus a total detuning of around -2 MHz was expected when the coolant inlet temperature is 15 degrees above ambient temperature and the cavity is fed with 350 W average power. 
The cavity incorporated two parallel circuits of 5.5~mm diameter sized to cool down 350~W (power corresponding to 260~MV/m peak surface electric field) with a 2.5~l/min water flow per circuit in turbulent regime.

The prototype is made of UNS C10100 OFE copper. It was machined at VECA s.r.l.(Italy) with a cell profile tolerance of 20~\textmu m and a surface roughness (Ra) of 0.4~\textmu m. The cavity elements were cleaned (degreasing, pickling and passivation) at CERN (Switzerland) and vacuum  brazed at Bodycote (France). 

Once the prototype was ready, the cavity was tuned and matched. The tuning was done by deforming the nose cone region from both sides of the bore hole with a bar clamp, in order to reduce the gap length of the cell and hence to decrease the resonant frequency. Then, the cavity was matched by adjusting the short plate position. After this procedure, a quality factor within 5\% of simulation estimations and a reflection coefficient of -27~dB were measured. Table~\ref{tab:mainEM} shows the main electromagnetic quantities of the test cavity after tuning and matching.  

\begin{table}[ht] 
	\centering
	\caption{Main 3 GHz test cavity quantities.}
		\begin{tabular}{l c c}
		  \hline
		  \textbf{Electromagnetic quantities} & \\
		  \hline
		  Frequency, $f_{0}$ & $GHz$ & 3 \\
		  Quality factor, $Q_{0}$ & $--$ & 9140\\
		  Coupling factor, $\beta$ & $--$ & 0.92 \\
		  $E_{max}/E_{0}$ & $--$ & 6.5 \\
		  $H_{max}/E_{0}$ & $kA/MV$ & 2.96 \\
		  $\sqrt{S_{c,max}}/E_{0}$ & $\frac{\sqrt{MW/mm^{2}}}{MV/m}$ & 0.032\\
		  \hline
		  \textbf{For operation at $E_{max}$=150 MV/m} & \\
		  \hline
		  \hline
		  Required Power, $P$ & $kW$ & $\sim$ 128 \\
		  Pulsed Surface Heating, $\Delta T$ & $degrees$ & 2 \\ % (for $t_{pulse}$=4~{\textmu}s)
		  \hline
			\end{tabular}
	\label{tab:mainEM}
\end{table} %\end{table*} 

\section{High-power tests in the CLIC Test Facility}
The primary goal of the high-power tests was to evaluate, preferably with a direct measurement, the breakdown rate at field levels in the range of the operation field of CABOTO. The $E_{max}/E_{0}$ for CABOTO is 4--5. For an accelerating gradient of 35~MV/m, the maximum electric field reached in the structure is about 170~MV/m, equivalent to the value that should be reached in the test cavity to perform this direct measurement.    

\begin{figure}[ht]
\begin{center}
\includegraphics[width=9cm]{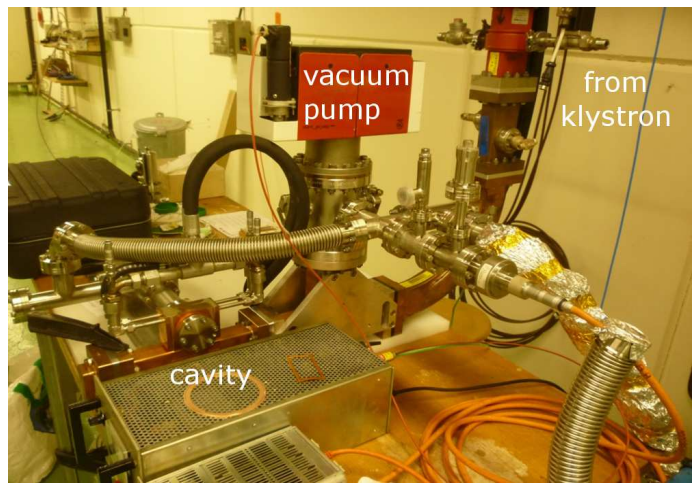}
\end{center}
\caption{\label{fig:expSetup} Experimental setup in the CLIC Test Facility.}
\end{figure}

The 3 GHz single-cell cavity underwent three high-power tests, conducted in the CLIC Test Facility (CTF) at CERN in February 2010, September-October 2011 and February-March 2012 (see Fig.~\ref{fig:expSetup}). The cavity saw about $5\times10^{7}$ RF pulses and suffered more than $10^{4}$ RF breakdowns during conditioning.  

The diagnostics included measurements of Faraday cup signals of field-emission currents (dark currents), incident and reflected RF power signals (amplitude and phase), thermocouples for temperature measurement of the cavity outer surface and the coolant, and a photomultiplier to detect light emitted during RF breakdowns.  

\section{Measurements and results} 

The breakdown rate, given in breakdowns per pulse per meter, was calculated as the number of breakdowns $N_{bd}$ over the number of RF pulses $N_{pulses}$ and the total length of the cell, $L_{cell}$=0.0189 m (see Eq.~\ref{eq:BDR_definition}). Breakdowns were identified from the increase in reflected power within an RF pulse. 

\begin{equation}
BDR\left[bpp/m\right] = \frac{N_{bd}\left[breakdowns\right]}{N_{pulses}\left[pulses\right]L_{cell}\left[m\right]}
\label{eq:BDR_definition}
\end{equation}
 
Fig.~\ref{fig:grudiev} shows the breakdown rate measurements for different electric field values. The lowest field measurement, at $E_{max} \sim$ 170~MV/m, is just a maximum threshold to the breakdown rate corresponding to that field range, as it was mentioned above.

\subsection{Consequences for TERA applications}
The maximum BDR threshold measured for $E_{max}$ of 170~MV/m (which corresponds to a maximum accelerating gradient $E_{0}$ of about 35~MV/m in CABOTO) and a RF pulse length of 2.5~\textmu s flat top (typical for CABOTO) is $3\times10^{-6}$ bpp/m. This breakdown rate value is acceptable for medical purposes. 

\subsection{Fitting models}

Two different models try to explain the appearance of RF breakdowns when operating at high gradient. The breakdown rate measurements performed for the 3 GHz test cavity were fitted to both models. 
 
\subsubsection{Power flow model}
The modified Poynting vector, a local field quantity that takes into account active and reactive power flow on the structure surfaces was proposed as a candidate to explain the high-gradient limit due to vacuum RF breakdowns~\cite{bib:Grudiev}. The field-emission currents at potential breakdown sites cause pulsed local heating of tips which acts as breakdown triggers in the model.

\begin{figure}[ht]
\begin{center}
\includegraphics[width=9cm]{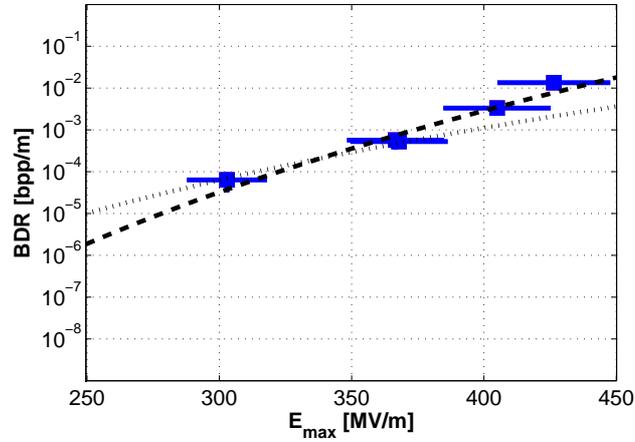}
\end{center}
\caption{\label{fig:grudiev} Breakdown rate measurements performed for the 3 GHz test cavity (blue squares). The blue arrow indicates that the point is a maximum threshold for the given electric field. The power flow model is fitted to low-field data, $E_{max}\in\left[300,370\right]$ MV/m (dashed line), and high-field data, $E_{max}\in\left[370,420\right]$ MV/m (dotted line).}
\end{figure}

The model says that the breakdown rate is proportional to the 15th power of the modified Poynting vector, $BDR \propto S_{c}^{15}$. Fig.~\ref{fig:grudiev} shows the breakdown rate measurements performed for the 3 GHz test cavity fitted with the power flow model. Data in the low-field range $E_{max}\in\left[300,370\right]$~MV/m fit to the model with a power of (10$\pm$5) for the electric field. Data in the high-field range $E_{max}\in\left[370,420\right]$ MV/m fit to the model with a power of (16$\pm$5) for the electric field. Both fits present the same root mean square deviation, $\epsilon_{RMS}=0.05$. 

\subsubsection{Stress model}
\begin{figure}[ht]
\begin{center}
\includegraphics[width=9cm]{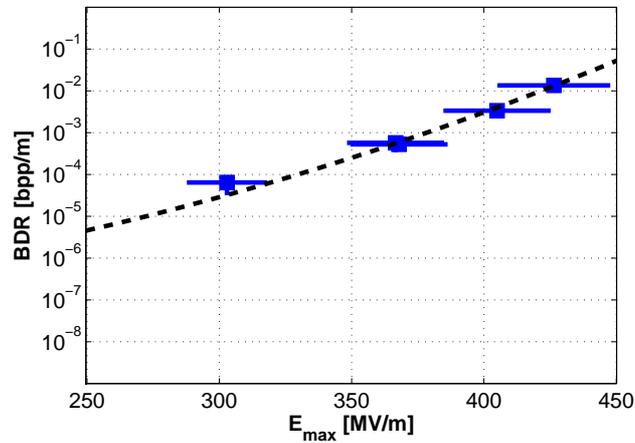}
\end{center}
\caption{\label{fig:stress} Breakdown rate measurements performed for the 3 GHz test cavity fitted to the stress model. The blue arrow indicates that the point is a maximum threshold for the given electric field.}
\end{figure}

A recently proposed model explains that limitation to high-gradient performance of accelerating structures due to RF breakdown comes from the stress caused on the crystalline structure of copper due to the exposure to an external electric field, which triggers the RF breakdown. The model assumes that the breakdown rate is proportional to the number of defects in the crystalline structure of copper~\cite{bib:Djurabekova}. The model follows the exponential growing fit:

\begin{equation}
BDR\left[bpp/m\right] \propto e^{\frac{\epsilon_{0}\Delta V}{kT} \left(E_{max}\left[MV/m\right]\right)^{2}}
\label{eq:stress}
\end{equation}

where $\epsilon_{0}$ is the permittivity in free space, $\Delta V$ is the defect volume, and $k$ is the Boltzmann constant. The dislocation loop radius $r_{loop}$ is calculated from $\Delta V$. Results from other experimental data lead to $\Delta V = \left[0.8, 13\right]\cdot10^{-25}$ $m^{3}$ and $r_{loop}=\left[13,40\right]$ $nm$. Fig.~\ref{fig:stress} shows the breakdown rate measurements performed for the 3 GHz test cavity fitted to the stress model with a root mean square deviation of $\epsilon_{RMS}=0.003$. The calculated values for $\Delta V$ and $r_{loop}$ are, respectively, $(13\pm6)\cdot 10^{-25}$ $m^{3}$ and $(46\pm11)$ $nm$, and are consistent with the values obtained for data from other experiments. 

\section{Conclusions and outlook}
The results of the 3~GHz cavity test are encouraging and show that structures can be operated with a RF flat top pulse length of 2.5~\textmu s and a surface electric field of $E_{max}=170$~MV/m (accelerating gradient $E_0=35$~MV/m) when limiting the break down rate to $10^{-6}$ per pulse per meter. Future high-power tests of the 5.7 GHz single-cell cavities will serve to understand which frequency would be more suitable for the construction of a medical high-gradient linac. The design, prototyping and high-power tests of a multicell structure operating at the selected frequency will follow. 

\ack
We would like to express our gratefulness to the CTF3 group for technical and scientific support to prepare and perform the experiment and to the CLIC RF structure development group for the enriching discussions about the preparation, development and analysis of the test.

%Thanks to the CERN PS group for the surveillance during the night shifts. We also acknowledge 
   
%We are also grateful to Rolf Wegner, who led the design, prototyping and first high-power test of the cavity, for the valuable discussions on the continuation of the high power tests. 

%Special thanks go to Alexey Dubrovskiy and Luca Timeo, for their special involvement in the experiment. Thanks also to Javier Bilbao de Mendizábal and Eugenio Bonomi. 
Special thanks go to Alexey Dubrovskiy and Luca Timeo, for their extensive contributions to the tuning and high power tests. Thanks also to Javier Bilbao de Mendizábal and Eugenio Bonomi. 

We are grateful to Vodafone Italy foundation for the fundings received to produce the test cavity. The research leading to this results was partially funded by the Seventh Framework Programme [FP7/2007-2013] under grant agreement no 215840-2.


\section*{References}
\begin{thebibliography}{9}
%\bibitem{LIBO} U. Amaldi et al., ``LIBO — a linac-booster for protontherapy: construction and tests of a prototype''. Nuclear Instruments and Methods in Physics Research A 521 (2004) 512–529. 
\bibitem{bib:Amaldi_NIMA10} U. Amaldi, R. Bonomi, S. Braccini, M. Crescenti, A. Degiovanni, M. Garlasché, A. Garonna, G. Magrin, C. Mellace, P. Pearce, G. Pittà, P. Puggioni, E. Rosso, S. Verd\'u-Andr\'es, R. Wegner, M. Weiss and R. Zennaro, ``Accelerators for hadrontherapy: From Lawrence cyclotrons to linacs''. Nuclear Instruments and Methods in Physics Research Section A: Accelerators, Spectrometers, Detectors and Associated Equipment 620 (2010) 563 - 577.
\bibitem{bib:Degiovanni_NIM11} A. Degiovanni, U. Amaldi, R. Bonomi, M. Garlasch\'e, A. Garonna, S. Verd\'u-Andr\'es and R. Wegner, ``TERA high gradient test program of RF cavities for medical linear accelerators''. Nuclear Instruments and Methods in Physics Research Section A: Accelerators, Spectrometers, Detectors and Associated Equipment 657 (2011) 55 - 58.
\bibitem{bib:Grudiev} A. Grudiev, S. Calatroni, and W. Wuensch, ``New local field quantity describing the high gradient limit of accelerating structures''. Phys. Rev. ST Accel. Beams 12, 102001 (2009).
%\bibitem{Alesini} D. Alesini et al., ``High Gradient Test of a C-Band Accelerating Structure Prototype for Energy Upgrade of Frascati Photoinjector SPARC'', Technical Note SPARC-RF-11/005 and KEK Report 2011-6 (2011).
\bibitem{bib:Djurabekova} F. Djurabekova et al., ``Multiscale modeling of electrical breakdown at high electric fields''. Talk in the International Workshop on Mechanisms of Vacuum Arcs MeVArc, Helsinki, Finland (2011).
\end{thebibliography}
\end{document}